\documentclass[12pt]{article} 
\usepackage{amsmath} 
\usepackage{amssymb} 
\usepackage{graphicx} 
\usepackage[dvips]{color} 
\usepackage{colordvi} 

\newcommand\fverb{\setbox\pippobox=\hbox\bgroup\verb}
\newcommand\fverbdo{\egroup\medskip\noindent%
			\fbox{\unhbox\pippobox}\ }
\newcommand\fverbit{\egroup\item[\fbox{\unhbox\pippobox}]}
\newbox\pippobox

\begin{document}
\thispagestyle{empty} 
\addtocounter{page}{-1} 
\vskip-0.35cm 
\begin{flushright} 
TIFR/TH/09-42\\ 
\end{flushright} 
\vspace*{0.2cm} 

\begin{center}
{\Large \bf Renormalization group flows in a} \\
{\Large \bf Lifshitz-like four fermi model}
\end{center}

\vspace*{0.2cm} 
\vspace*{1.0cm}
\centerline{\bf Avinash Dhar, Gautam Mandal and Partha Nag} 
\vspace*{0.7cm} 
\centerline{\it Department of Theoretical Physics} 
\vspace*{0.2cm} 
\centerline{\it Tata Institute of Fundamental Research,}  
\vspace*{0.2cm} 
\centerline{\it Mumbai 400 005, \rm INDIA} 

\vspace*{0.5cm} 
\centerline{\tt email: adhar, mandal, parthanag@theory.tifr.res.in} 
 
\vspace*{0.8cm} 
\centerline{\bf Abstract} 
\vspace*{0.3cm} 
\vspace*{0.5cm} 

We study renormalization group flows in the Lifshitz-like $N$-flavour
four fermi model discussed in 0905.2928. In the large-$N$ limit, a
nontrivial flow occurs in only one of all possible marginal couplings
and one relevant coupling, which provides the scale for Lorentz
invariance violations. We discuss in detail the phase diagram and RG
flows in the space of couplings, which includes the Lifshitz fixed
point, the free field fixed point and a new fixed point characterized
by $z=1$ scaling and a violation of Lorentz invariance, which cannot
be tuned away by adjusting a parameter. In the broken symmetry phase,
the model flows from the $z=3$ Lifshitz-like fixed point in the
ultraviolet to this new fixed point in the infrared. However, in a
modified version of the present model, which has an effective
ultraviolet cut-off much smaller than the Lorentz invariance violating
scale, the infrared behaviour is governed by an approximately Lorentz
invariant theory, similar to the low energy limit of the usual
relativistic Nambu$-$Jona-Lasinio model. Such a modified model could
be realized by a supersymmetric version of the present model.

\baselineskip=15pt 
  
\def\gap#1{\vspace{#1 ex}} 
\def\be{\begin{equation}} 
\def\ee{\end{equation}} 
\def\bal{\begin{array}{l}} 
\def\ba#1{\begin{array}{#1}}  
\def\ea{\end{array}} 
\def\bea{\begin{eqnarray}} 
\def\eea{\end{eqnarray}} 
\def\beas{\begin{eqnarray*}} 
\def\eeas{\end{eqnarray*}} 
\def\del{\partial} 
\def\eq#1{(\ref{#1})} 
\def\fig#1{Fig \ref{#1}}  
\def\re#1{{\bf #1}} 
\def\bull{$\bullet$} 
\def\mm{&&\kern-18pt}  
\def\nn{\\\nonumber} 
\def\ub{\underbar} 
\def\nl{\hfill\break} 
\def\ni{\noindent} 
\def\bibi{\bibitem} 
\def\ket{\rangle} 
\def\bra{\langle} 
\def\vev#1{\langle #1 \rangle}  
\def\lsim{\stackrel{<}{\sim}} 
\def\gsim{\stackrel{>}{\sim}} 
\def\mygsim{\lower .7ex\hbox{$\gsim$}}
\def\mattwo#1#2#3#4{\left( 
\begin{array}{cc}#1&#2\\#3&#4\end{array}\right)}  
\def\tgen#1{T^{#1}} 
\def\half{\frac12} 
\def\floor#1{{\lfloor #1 \rfloor}} 
\def\ceil#1{{\lceil #1 \rceil}} 
\def\slash#1{{#1}\kern-8pt/\kern2pt}

 \def\mysec#1{\gap1\ni{\bf #1}\gap1} 
 
\def\bit{\begin{item}} 
\def\eit{\end{item}} 
\def\benu{\begin{enumerate}} 
\def\eenu{\end{enumerate}} 
\def\a{\alpha} 
\def\as{\asymp} 
\def\ap{\approx} 
\def\b{\beta} 
\def\bp{\bar{\partial}} 
\def\cA{{\cal{A}}} 
\def\cD{{\cal{D}}} 
\def\cL{{\cal{L}}} 
\def\cP{{\cal{P}}} 
\def\cR{{\cal{R}}} 
\def\da{\dagger} 
\def\de{\delta} 
\def\e{\eta} 
\def\ep{\epsilon} 
\def\eqv{\equiv} 
\def\f{\frac} 
\def\g{\gamma} 
\def\G{\Gamma} 
\def\h{\hat} 
\def\hs{\hspace} 
\def\i{\iota} 
\def\k{\kappa} 
\def\lf{\left} 
\def\l{\lambda} 
\def\la{\leftarrow} 
\def\La{\Leftarrow} 
\def\Lla{\Longleftarrow} 
\def\Lra{\Longrightarrow} 
\def\L{\Lambda} 
\def\m{\mu} 
\def\na{\nabla} 
\def\nn{\nonumber\\} 
\def\om{\omega} 
\def\O{\Omega} 
\def\p{\phi}
\def\tp{{\tilde \pi}}
\def\P{\Phi} 
\def\pa{\partial} 
\def\pr{\prime} 
\def\r{\rho} 
\def\ra{\rightarrow} 
\def\Ra{\Rightarrow} 
\def\ri{\right} 
\def\s{\sigma} 
\def\ts{{\tilde \sigma}}
\def\sq{\sqrt} 
\def\S{\Sigma} 
\def\si{\simeq} 
\def\st{\star} 
\def\t{\theta} 
\def\ta{\tau} 
\def\ti{\tilde} 
\def\tm{\times} 
\def\tr{\textrm} 
\def\T{\Theta} 
\def\up{\upsilon} 
\def\Up{\Upsilon} 
\def\v{\varepsilon} 
\def\vh{\varpi} 
\def\vk{\vec{k}} 
\def\vp{\varphi} 
\def\vr{\varrho} 
\def\vs{\varsigma} 
\def\vt{\vartheta} 
\def\w{\wedge} 
\def\z{\zeta} 
\def\psd{{\psi^\dagger}}
\def\Psd{{\Psi^\dagger}}
\newpage

\tableofcontents 

\title{Lifshitz}

\section{Introduction}

Recently, there has been considerable interest in field theories in
which Lorentz invariance is explicitly violated by terms containing
higher order spatial derivatives. The presence of these terms
leads to softer ultraviolet behaviour while preserving unitarity
(which is typically lost in the presence of Lorentz invariant higher
derivative terms because of ghosts associated with higher time
derivatives). At very high energies, these Lorentz invariance
violating terms dominate, leading to Lifshitz-like anisotropic scaling
symmetry (in the classical theory) in which time and space scale
differently: $x \to x/a, t \to t/a^z$. The exponent $z$ 
characterizes the scaling symmetry.

Lifshitz-like field theories with anisotropic scaling have been used
in condensed matter systems to describe quantum criticality
\cite{Hornreich:1975zz}-\cite{Son:2007ja}. Recently they have 
also been discussed in string theory in the context of possible
applications of AdS/CFT duality 
\cite{Son:2008ye}-\cite{Hartnoll:2009sz} to
condensed matter systems involving strongly interacting
constituents. In a separate development, the idea that a relativistic
theory at low energies may have a Lorentz non-invariant ultraviolet
completion was suggested in \cite{Horava:2008jf}. This idea has been
further explored in \cite{Visser:2009fg}-\cite{Chao:2009dw}. The
suggestion that an ultraviolet completion of quantum gravity may be
similarly formulated \cite{Horava:2009uw} has serious difficulties
\cite{Charmousis:2009tc}-\cite{Papazoglou:2009fj} because gravity elevates 
Lorentz invariance to a local gauge symmetry, which cannot be broken
except by some kind of Higgs mechanism. In the present work we will
focus only on non-gravitational theories.

Lifshitz-like field theories with Lorentz invariance violations (LIV)
have also recently been discussed in the context of applications to
particle physics
\cite{Anselmi:2008bt,Anselmi:2009vz,Dhar:2009dx,Chao:2009dw,Kawamura:2009re,Kaneta:2009ci}. In
\cite{Anselmi:2009vz,Dhar:2009dx} it was argued that a $z=3$
Lifshitz-like ultraviolet completion of the Nambu$-$Jona-Lasinio (NJL) model
\cite{Nambu:1961tp} in $3+1$ space-time dimensions has the required
properties to replace the Higgs sector of the electro-weak theory. The
four-fermion coupling in this model is asymptotically free, leading to
dynamical mass generation for the fermions and chiral symmetry
breaking. In an appropriately gauged version, fluctuations of the
magnitude of the fermion bilinear order parameter\footnote{The phase
  of the fermion bilinear is the Goldstone mode which combines with
  the gauge field as usual to make it massive.} can be interpreted as
the Higgs field. This model obviates the need for fine tuning at the
expense of introducing LIV at high energies. The hope is that for a
sufficiently large LIV scale, the low-energy theory could be
consistent with experimental constraints on LIV\footnote{For current
  situation on experimental searches of Lorentz symmetry violations,
  see \cite{Liberati:2009pf}-\cite{Scully:2008jp}.}. This may,
however, require a new fine tuning of parameters \cite{Iengo:2009ix}.

The main purpose of this work is to analyze renormalization group (RG)
flows in the model of \cite{Dhar:2009dx} to understand in detail
the possible emergence of Lorentz invariance at low energies. The
analysis has been performed in the leading large-$N$
approximation. Our findings can be summarized as follows.

\medskip

\noindent $\bullet$ At low energies, in the fermionic sector, the
theory recovers approximate Lorentz invariance, violations being of
order $E^2/\mu^2$, where $\mu$ is the energy scale associated with
LIV. However, in the bosonic sector, in the broken chiral symmetry
phase, the induced kinetic terms violate Lorentz invariance at
${\cal{O}}(1)$ level. The origin of these violations is simple to
understand $-$ they arise from fermion modes with energies higher than
$\mu$ propagating in a loop.  These violations can be made small by
imposing an effective cut-off on the theory and arranging $\mu$ to be
much larger than the cut-off (see the last paragraph of Section
\ref{sec-low}).

\medskip

\noindent $\bullet$ The RG flows reveal a new nontrivial fixed point,
apart from the $z=3$ ultraviolet fixed point. The theory flows down
from high energies to this fixed point at low energies. The above
mentioned approximate Lorentz invariance in the fermionic sector and
${\cal{O}}(1)$ LIV in the bosonic sector are characteristic of this
new fixed point. If one works with a fixed finite cut-off $\L$, and
sends the LIV scale $\mu$ much above $\L$, the Lorentz violations
become smaller and smaller, leading to an approximately Lorentz
invariant theory which is identical to an effective theory derived
from the NJL model at low energies, with a cut-off $\L$.

\medskip

The plan of the paper is as follows. In section 2 we deform the z=3
action with all possible marginal and relevant couplings (from the
point of view of $z=3$ scaling) and study their effect on the vacuum
solution. We find that only one of the three possible four-fermion
(marginal) couplings has a nontrivial flow. Moreover, there is only
one relevant coupling that affects the low energy physics, namely the
coupling that determines the scale of LIV. This had already been
remarked in \cite{Dhar:2009dx}, but section 2 provides a detailed
justification for it. In section 3, we derive the effective action at
a scale much smaller than the scale of LIV. We find that at low
energies, in the broken symmetry phase, the kinetic terms induced by
fermion loops for the massive scalar bound state of the fermions
violate Lorentz invariance at ${\cal{O}}(1)$ level, which cannot be
corrected by any fine-tuning of parameters. In section 4 we study the
RG flow of the couplings and locate fixed points. We compare this
fixed point structure with that in the relativistic NJL model.  We end
with some concluding remarks in section 5. Details of some of the
calculations have been given in three Appendices.

\section{The action and vacuum solutions}

The model discussed in \cite{Dhar:2009dx} consists of 2N species of
fermions, $\psi_{ai}(t,\vec x)$, where the index $a$ runs over the
values $1,~2$ and the index $i$ runs from 1 to $N$. Each of these
fermions is an $SU(2)$ spinor, where $SU(2)$ is the double cover of
the spatial rotation group $SO(3)$. It is useful to view the index $a$
as denoting the two Weyl components of a Dirac fermion in a four
dimensional theory with Lorentz invariance. The action we consider
will have the following symmetries: a global $U(1)_1 \times U(1)_2$
symmetry\footnote{This is the analogue of chiral symmetry in the
corresponding Lorentz invariant four fermi model.} under which the
fermions transform as
\bea
 \psi_{ai} \to e^{i\alpha_a} \psi_{ai},\hspace{0.5cm} a=1,2;
\label{chiral}
\eea
and a global $U(N)$ symmetry under which the fermions transform in the
fundamental representation
\bea 
\psi_{ai} \to U_{ij}\psi_{aj}, \hspace{0.5cm} i=1,...,N.
\eea
In addition to these symmetries, we will ensure that the action is
invariant under the interchange $\psi_{1i}(t, \vec x)
\to \psi_{2i}(t,-\vec x) $. This is the analogue of the parity operation 
in the relativistic Dirac theory.

A general action which is consistent with the above symmetries and
contains all the relevant and marginal couplings is given
by\footnote{All other possible four-fermion terms, like
e.g. $(\psi_{1i}^\dagger \psi_{2j})(\psi_{2j}^\dagger \psi_{1i})$, can
be related to the terms in \eq{relevant} and/or terms involving vector
bilinears like $(\psi_{1i}^\dagger \vec{\sigma}
\psi_{1i}).(\psi_{2j}^\dagger \vec{\sigma}\psi_{2j})$. The latter do 
not affect vacuum solutions since the vevs of such bilinears vanish
because of invariance under spatial rotations.}
\bea
\!\!\!S =\!\! \int\!\!\mm  d^3\vec  x\  dt \left[ 
 \psd_{\!1i}
\left\{i \del_t - i \vec \del. \vec \s\ \left(g_0(-i\vec{\del})^2+ g_1\right)
- g_2 \vec{\del}^2 \right\} \psi_{1i} 
\right.
\nn
\mm
+  \psd_{\!2i} 
\left\{i \del_t + i \vec \del. \vec \s\ \left(g_0(-i\vec{\del})^2+ g_1\right)
- g_2 \vec{\del}^2 \right\} \psi_{2i}
+  g_3 \left( \psd_{\!1i} \psi_{1i} +  
\psd_{\!2i} \psi_{2i} \right)  
\nn
\mm
\left.
 + g_4^2 \left\{\left(\psd_{\!\!1i}\psi_{1i}\right)^2 + 
\left(\psd_{\!\!2i} \psi_{2i}\right)^2\right\}
+ g_5^2 \left( \psd_{\!\!1i}\psi_{1i} \psd_{2j}\psi_{2j} \right)
+ g^2 \left( \psd_{\!\!1i}\psi_{2i} \psd_{2j}\psi_{1j} \right)
 \right].
\nn
\label{relevant}
\eea
At high energies, this action has the scale invariance of a $z=3$
Lifshitz-like theory in which space and time have the dimensions
$[x]=-1$ and $[t]=-3$. At low energies, the relevant term with
coupling $g_1$ dominates, so one expects the model to flow down to
an approximately Lorentz invariant theory at low energies. We will
study the dynamics of this action in the large $N$ limit. 

It is useful to employ the notation of Dirac matrices and rewrite the
action \eq{relevant} in terms of the four-component spinor
\bea
\Psi_i = \left(\begin{array}{c} \psi_{1i} \\ \psi_{2i}
\end{array} \right)
\label{dirac1}
\eea
and its Weyl components $P_{L,R}\Psi_i={\Psi_i}_{L,R}=\psi_{1,2i}$. 
We will also use the Dirac gamma matrices
\bea
\gamma^0 = \s_1 \otimes {\bf 1}, \quad \g^i = i\s_2 \otimes \s_i, \quad
\g^5 =i\g^0 \g^1 \g^2 \g^3 = \s_3 \otimes {\bf 1},
\label{dirac2}
\eea
where the $\sigma$'s are standard Pauli matrices satisfying
$[\sigma_i,\sigma_j]=i\epsilon_{ijk}\sigma_k$ and ${\bf 1}$ is the
$2\times 2$ identity matrix. In terms of these matrices, 
the Weyl projection operators are given by $P_{L,R}=\frac{1}{2}(1 \pm
\gamma^5)$.
In this notation, the above action takes the form
\begin{eqnarray}
S=\int  \mm d^3\vec  x\  dt  \biggl[ \bar{\Psi}_i \biggl\{\gamma^0 
\left(i\partial_0-g_2 \vec{\partial}^2+g_3\right)
+i \vec{\gamma}.\vec{\partial} \left(g_1-g_0\vec{\partial}^2\right)
\biggr\} \Psi_i \nonumber\\
\mm +g_4^2 \biggl\{(\bar{\Psi}_{Li} \gamma^0 \Psi_{Li})^2+
(\bar{\Psi}_{Ri} \gamma^0 \Psi_{Ri})^2\biggr\}+
g_5^2(\bar{\Psi}_{Li} \gamma^0 \Psi_{Li})(\bar{\Psi}_{Rj} \gamma^0 \Psi_{Rj}) 
\nonumber\\ \mm 
+g^2 (\bar{\Psi}_{Li}\Psi_{Rj})(\bar{\Psi}_{Rj}\Psi_{Li})
\biggr],
\label{actiondirac}
\end{eqnarray}

As is usual for actions with four-fermion interactions, we now
introduce auxiliary scalar fields to rewrite the above in the
completely equivalent form of an action quadratic in fermions:
\bea
S=\int  \mm d^3\vec  x\  dt  \biggl[ \bar{\Psi}_i \biggl\{\gamma^0 
\left(i\partial_0-g_2 \vec{\partial}^2+g_3\right)
+i \vec{\gamma}.\vec{\partial} \left(g_1-g_0\vec{\partial}^2\right)
+\left(\phi P_L +\phi^{\ast} P_R\right)\nonumber\\
\mm+\gamma^0 \left(\alpha P_L+\beta P_R\right)\biggr\} \Psi_i 
-\frac{\rho^2+\rho\eta}{g_4^2}-\frac{g_4^2\eta^2}{g_5^2}
-\frac{|\phi|^2}{g^2}\biggr].
\label{actionboson}
\eea
Here $\rho$ and $\eta$ are real scalar fields and $\phi$ is a
complex scalar field. We have also defined 
\bea
2\rho+\eta \equiv \alpha, \qquad 
\frac{g_5^2}{g_4^2}\rho +2\frac{g_4^2}{g_5^2}\eta \equiv \beta.
\label{scalars}
\eea
The scalars are fermion bilinear composites, as can be 
easily derived from their equations of motion:  
\bea
\rho=g_4^2(\bar{\Psi}_{Li} \gamma^0 \Psi_{Li}), \qquad
\eta=g_5^2(\bar{\Psi}_{Ri} \gamma^0 \Psi_{Ri}), \qquad
\phi=g^2(\bar{\Psi}_{Li} \Psi_{Ri}).
\label{bilinears}
\eea

\subsection{Vacuum solutions}

Since the action \eq{actionboson} is quadratic in fermions, one can
integrate these out to get an effective action for the scalar
fields. For vacuum solutions, it is sufficient to consider only the
homogeneous modes, $\rho_0,~\eta_0$ and $\phi_0$, of the scalar
fields. Moreover, without loss of generality, one may take $\phi_0$ to
be real. In this case, the effective action for the scalars is given
by
\begin{eqnarray}
S_{eff}=-i \mm  N \ {\rm Tr} \ln \biggl\{l_0\gamma^0-\vec{l}.\vec{\gamma}
+\phi_0+\gamma^0(\alpha_0 P_L+\beta_0 P_R)\biggr\}\nonumber\\
\mm-\left(\frac{\rho_0^2+\rho_0\eta_0}{g_4^2}+\frac{g_4^2\eta_0^2}{g_5^2}
+\frac{\phi_0^2}{g^2}\right) V,
\label{effective action}
\end{eqnarray}
where $V$ denotes the volume of space-time. Also, in the momentum space 
representation, we have
\begin{eqnarray}
l_0(k_0,\vec{k})=k_0+g_2 k^2+g_3, \qquad
\vec{l}(\vec{k})=\vec{k}(g_0k^2+g_1), \qquad 
k=|\vec{k}|.
\label{l}
\end{eqnarray}

The equations of motion are obtained by varying this action with
respect to $\rho_0,~\eta_0$ and $\phi_0$. The calculation can be
simplified by noticing that due to the ``parity'' symmetry $\Psi_{Li}(t,
\vec x) \to \Psi_{Ri}(t,-\vec x)$, which we assume is unbroken, 
and the relations \eq{bilinears}, the vacuum solutions $\rho_0$ and
$\eta_0$ are related, i.e.
\bea
\rho_0/g_4^2=\eta_0/g_5^2.
\label{vacuumrel} 
\eea
Thus, there are really only two independent variables, $\rho_0$ and
$\phi_0$ . Using this relation after varying \eq{effective action}
with respect to $\rho_0,~\eta_0$ and $\phi_0$, the two independent
equations of motion can be written as
\bea
&&\frac{\rho_0}{\lambda_4}=-2i \int \frac{d^4k}{(2\pi)^4} 
\frac{(l_0+\alpha_0)}{(l_0+\alpha_0)^2-\vec{l}^2-\phi_0^2}
\label{rho},\\ 
&&\frac{1}{\lambda}=2i\int \frac{d^4k}{(2\pi)^4} 
\frac{1}{(l_0+\alpha_0)^2-\vec{l}^2-\phi_0^2}.\label{phi} 
\eea
where we have defined the 't Hooft couplings, 
\bea
g_{4,5}^2N \equiv \lambda_{4,5}, \qquad g^2N \equiv \lambda,
\label{thooft}
\eea
and made use of the relation 
\bea
\alpha_0=\beta_0=\left(2+\frac{\lambda_5}{\lambda_4}\right)\rho_0,
\label{alphabeta}
\eea 
which follows from equations \eq{scalars} and \eq{vacuumrel}. The
equations \eq{rho} and \eq{phi} are exact in the usual large N limit
in which the 't Hooft couplings are held fixed.

Let us first consider equation \eq{rho}. The momentum integral on the
right hand side has a power divergence, so the dependence on
$\alpha_0$ cannot be removed by shifting $k_0$. To do the integral, we
need to introduce a regulator. We will use a simple cut-off
regulator\footnote{Note that the regulator used here explicitly breaks
Lorentz invariance. This does not affect vacuum solutions. However, in
sections 3 and 4, where we consider the question of restoration of
Lorentz invariance at low energies, it will be important to choose a
suitably different regulator that allows for such a
possibility.}. Continuing to Euclidean momenta, we may then write
\bea
\tilde{\rho}_0 &=& \frac{\lambda_4}{2\pi^3} 
\int_{0}^{M} dk \ k^2 \int_{-M^3}^{M^3} dk_0 \ 
\frac{k_0-g(k)}{\left(k_0-g(k)\right)^2+l^{2}+\phi_0^2}
\nonumber \\
&=& \frac{\lambda_4}{4\pi^3}\int_{0}^{M} dk \ k^2 \ \ln  
\left \{\frac{l^{2}+\phi_0^2+\left(M^3-g(k)\right)^2}
{l^{2}+\phi_0^2+\left(M^3+g(k)\right)^2}\right\},
\label{rhointegral}
\eea
where $l=|\vec{l}|=k|g_0k^2+g_1|$,
$g(k)=(\tilde{g}_2k^2+\tilde{g}_3+\tilde{\alpha_0})$ and
$\tilde{\rho}_0=i\rho_0$, $\tilde{g}_2=i g_2$ and $\tilde{g}_3=i g_3$
are Euclidean continuations of respectively $\rho_0$, $g_2$ and
$g_3$. 

In the limit of large $M$, the integral in \eq{rhointegral} can
be done by expanding both the numerator and denominator of the
argument of the `ln' function around $(M^6+g_0^2k^6)$. Discarding
terms that vanish as $M \rightarrow \infty$, we get
\begin{eqnarray}
\rho_0=
\frac{\lambda_4}{\pi^3}
\biggl\{-M^2 g_2 I_0-(g_3+\alpha_0) I_1
+\left(2 g_0 g_1-g_2^2\right)g_2 I_2
+\frac{4}{3} g_2^3 I_3\biggr\},
\label{rho-0}
\end{eqnarray}
where the coefficients $I_0,~I_1,~I_2$ and $I_3$ are functions of
$g_0$ only and are listed in Appendix 1.  In writing the above, we
have continued the various parameters and fields back to Minkowski
signature.  Now, using relation
\eq{alphabeta}, the above equation can be solved for $\rho_0$:
\begin{eqnarray}
&& \rho_0 \left\{1 + \left(2\lambda_4+\lambda_5 \right)\frac{I_1}{\pi^3} 
\right\} \nonumber \\
&& =\frac{\lambda_4}{\pi^3}\left\{-(g_3+M^2 g_2I_0/I_1)I_1
+\left(2 g_0 g_1-g_2^2\right)g_2 I_2+\frac{4}{3} g_2^3 I_3 \right\}.
\label{rhosolution}
\end{eqnarray} 

Notice that $\rho_0$ vanishes if $g_2=g_3=0$. If $g_2=0$ but $g_3 \neq 0$,
then $\rho_0$ is nonzero, but we may take all the couplings appearing
in the solution as fixed, independent of the cut-off. In case $g_2
\neq 0$, then the right hand side of \eq{rhosolution} diverges
quadratically with the cut-off. However, this divergence can be
removed by shifting $g_3$, i.e. we set $g_3=(g_3'-M^2
g_2I_0/I_1)$, where $g_3'$ is independent of the cut-off
$M$. This determines $\rho_0$ in terms of the
$M$-independent couplings $g_2$, $g_3'$, $g_4$, etc.

The other vacuum parameter, $\phi_0$, is determined by equation
\eq{phi}. In this case, the integral over $k_0$ on the right hand side 
is convergent and the entire integral is only logarithmically
divergent. So a shift of the integration variable $k_0$ is
allowed. Doing this enables us to get rid of the couplings
$g_2,~g_3,~g_4,~g_5$ and $\rho_0$ from this gap equation. After a
Euclidean continuation ($k_0=ik_4$) it then takes the form
\bea
\frac{1}{\lambda}=2\int \frac{d^4k}{(2\pi)^4} 
\frac{1}{(k_4^2+l^2(k)+\phi_0^2)}.
\label{gapeqn} 
\eea
As discussed in \cite{Dhar:2009dx}, this equation signals the
breaking of the global $U(1)_1 \times U(1)_2$ ``chiral'' symmetry
\eq{chiral} down to $U(1)$. It also requires that the coupling 
$\lambda$ have a nontrivial RG flow, which depends on the
couplings $g_0$ and $g_1$. We will analyze this flow in detail in
section 4.

We end this section with the following comment. It should be clear
from the above discussion that in the present model all the
interesting dynamics arises from the NJL type of four fermi
interaction (conjugate to the 't Hooft coupling $\lambda$) and its RG
flow is decoupled from the dynamics of all the other marginal
couplings. Therefore, for calculational simplicity, in the rest of the
paper we will set the marginal couplings $g_4$ and $g_5$ to zero. We
will also set $g_2=g_3=0$ since these couplings also do not affect the
RG flow of $\lambda$.   

\section{\label{sec-low} Low energy effective action}

Setting $g_2=g_3=g_4=g_5=0$ in \eq{actiondirac} as discussed
above, and going over to the
bosonic variable $\phi(x)$ as in \eq{actionboson}, we get the action 
\bea S= \int d^4x
\biggl[\mm\bar{\Psi}_i\biggl\{i\gamma^0\partial_0+i\vec{\gamma}.
  \vec{\partial}\left(g_1-g_{0} \vec{\partial}^2\right)+(\phi P_L
  +\phi^* P_R)
  \biggr\}\Psi_i-\frac{N}{\lambda}|\phi|^2\biggr]. \nonumber \\
\label{simpleaction}
\eea
We are interested in the low energy effective action for this system
in the phase in which the symmetry \eq{chiral} is broken. To derive
the low energy action, we substitute in the above
$\phi(x)=(\phi_0+\frac{\sigma(x)}{\sqrt{N}})
e^{i\frac{\theta(x)}{\sqrt{N}}}$, where $\phi_0$ is the vacuum
solution and $\sigma(x)$ and $\theta(x)$ are respectively the
magnitude and phase of the fluctuation of $\phi(x)$ around the vacuum.
We get,
\bea
S= \int d^4x \biggl[\mm\bar{\Psi}_i\biggl\{i\gamma^0\partial_0+i\vec{\gamma}.
\vec{\partial}\left(g_1-g_{0} \vec{\partial}^2\right)+\biggl(\phi_0+
\frac{\sigma}{\sqrt{N}}\biggr) 
\biggl(P_L e^{i\frac{\theta}{\sqrt{N}}}+
P_R e^{-i\frac{\theta}{\sqrt{N}}}\biggr) 
\biggr\}\Psi_i \nonumber\\
\mm -\frac{1}{\lambda} \sigma^2-\frac{2\sqrt{N}}
{\lambda}\phi_0 \sigma-\frac{N}{\lambda}\phi_0^2\biggr]. 
\label{simpleraction}
\eea
The phase $\theta$ is the Goldstone mode\footnote{If the $U(1)_1
\times U(1)_2$ symmetry is gauged (in such a manner that there are no
anomalies), then this mode will be eaten up by the gauge fields by the
usual Higgs mechanism.}.  It can be ``rotated'' away from the
Yukawa-type coupling by making the change of variables $\Psi_{iL}
\rightarrow e^{-i\frac{\theta}{2\sqrt{N}}}
\Psi_{iL}$ and $\Psi_{iR} \rightarrow e^{i\frac{\theta}{2\sqrt{N}}}
\Psi_{iR}$. Its derivatives will, however, now appear from the fermion
kinetic terms.  After this change of
variables, the action takes the form 
\bea 
S= \int d^4x
\biggl[\mm\bar{\Psi}_i\biggl\{i\gamma^0\partial_0+i\vec{\gamma}.
  \vec{\partial}\left(g_1 -g_{0} \vec{\partial}^2\right)+\phi_0+
  \frac{\sigma}{\sqrt{N}} \biggr\}\Psi_i+{\cal O}(\bar{\Psi}
  \del\theta \Psi) \nonumber \\ \mm -\frac{1}{\lambda} \sigma^2
  -\frac{2\sqrt{N}} {\lambda}\phi_0
  \sigma-\frac{N}{\lambda}\phi_0^2\biggr],
\label{simplestaction}
\eea 
where ${\cal O}(\bar{\Psi} \del \theta \Psi)$ indicates terms
involving derivatives of the Goldstone mode $\theta$. In the following
we will ignore these terms and retain only the mode $\sigma$. There is
no basic difficulty in retaining $\theta$ also, but this is not
essential for describing the RG flow of $\lambda$ and LIV properties
of the low energy action\footnote{There are no terms in the low energy
action which mix $\theta$ and $\sigma$ at the quadratic
level. Moreover, the induced kinetic terms for both the fields are
finite in the limit of infinite UV cut-off.  Only the coefficient of
the $\sigma^2$ term diverges and needs to be renormalized.  Hence, for
simplicity, in the following we have set $\theta$ to a constant.}.

To proceed further, it is convenient to rewrite the action
\eq{simplestaction} in momentum space. We have
\begin{eqnarray}
S = \mm\int \frac{d^4k}{(2\pi)^4} \int 
\frac{d^4q}{(2\pi)^4} \bar\Psi_i(k) 
\biggl\{\left(\gamma^0k_0-\vec{\gamma}.\vec{l}(\vec{k})+\phi_0\right)(2\pi)^4
\delta^4(k-q)
+\frac{1}{\sqrt{N}}\sigma(k-q)\biggr\}\Psi_i(q)\nonumber\\
\mm -\int \frac{d^4k}{(2\pi)^4} \frac{1}{\lambda} \ |\sigma(k)|^2 
-\frac{2\sqrt{N}}{\lambda}\phi_0 \sigma(k=0)-\frac{VN}{\lambda}\phi_0^2,
\label{momaction}
\end{eqnarray}
where $V$ denotes the volume of space-time and $\vec{l}(\vec{k})$ is
given by \eq{l}. As discussed in the next section, although this
action depends on the two parameters $g_0$ and $g_1$, the physics
described by it actually depend on only a particular combination of
these, namely
\bea
\mu=|g_1^3/g_0|^{\frac{1}{2}}.
\label{masslv}
\eea 
$\mu$ has the dimensions of energy. As discussed below, its
significance is that it is the energy scale at which Lorentz
violations become important. The action
\eq{momaction} can, in fact, be explicitly written in terms of this 
combination. However, two different versions of the action are needed
to cover the entire line of real values of $\mu$, from zero to
infinity. In the form \eq{momaction}, the action is valid for all
values, but at the expense of having a redundant variable. We will
discuss this issue further in the next section after we have obtained
the low energy action.

The action at low energies is obtained by integrating out high energy
modes from this action. To carry out this procedure, we will need to
regularize divergences in loop momentum integrals, which we will do by
using a cut-off that scales like energy. However, the way in which the
cut-off is imposed needs to be chosen carefully, because an
arbitrarily chosen cut-off procedure (in a theory with LIV) might
preclude the possibility of an approximate restoration of Lorentz
invariance at low energies. We choose to impose a cut-off on the
expression $(k_4^2+l^2(k))$, where $k_4=-ik_0$ is the Euclidean
continuation of energy.  To understand why it is natural to adopt this
cut-off procedure, recall the definition of
$l(k)=|\vec{l}(\vec{k})|=k|g_0k^2+g_1|$, given in \eq{l}. Now, for
$g_1 \neq 0$, we can completely scale it out of this expression by the
scaling $k \rightarrow k/|g_1|$. This gives $l \rightarrow
k|\epsilon \mu^{-2}k^2+1|$, where $\epsilon=\pm 1$ is the relative sign of
$g_0$ with respect to $g_1$. For $k << \mu$, $(k_4^2+l^2(k))$ can be
well approximated by the $SO(4)$-invariant form $(k_4^2+k^2)$. Thus,
the proposed cut-off is naturally consistent with the possibility of
Lorentz symmetry restoration at low energies\footnote{By contrast, a
cut-off such as the one used in \eq{rhointegral} is intrinsically
Lorentz non-invariant. With such a cut-off, possible emergence of
Lorentz invariance at low energies will not be manifest.}.

In the following we will use the notation $\int [d^4k]_{E \rightarrow \L}$ 
to indicate that the integral over Euclidean momenta is restricted
to the region $E^2 \leq (k_4^2+l^2(k)) \leq \L^2$, i.e. 
\bea
&& \int [d^4k]_{E \rightarrow \L} \ \cdots \ \equiv \nonumber \\
&&~~~~~~~\int d^4k \ \theta\left(\L^2-(k_4^2+l^2(k))\right)
\theta\left((k_4^2+l^2(k))-E^2\right) \ \cdots, 
\eea
where the $\theta$ is the usual step function. Note that both $\L$ and
$E$ have dimensions of energy. For the special case of $E=0$, it will
be convenient to use the more compact notation $\int [d^4k]_\L$.

Let us now integrate out from \eq{momaction} all the modes between
$(k_4^2+l^2(k))=\L^2$ and $(k_4^2+l^2(k))=E^2$, where $\L$ is the
cut-off and $E < \L$. The action with the lower energy cut-off $E$ is
given by
\begin{eqnarray}
S\kern-10pt &=& \kern-5pt 
\int \frac{[d^4k]_E}{(2\pi)^4} 
\kern-5pt \int \frac{[d^4q]_E}{(2\pi)^4} \
\bar\Psi_i(k) \biggl\{\kern-2pt
(2\pi)^4 \biggl(k_0 \gamma^0-
\vec{l}(\vec{k}).\vec{\gamma}
+\phi_0\biggr)\delta^4(k-q)+\frac{\sigma(k-q)}{\sqrt{N}}
\kern-2pt\biggr\}
\Psi_i(q)\nonumber\\
&& -\int \frac{[d^4k]_E}{(2\pi)^4} 
\left(C_0+C_1 k_0^2-C_2 k^2 \right)|\sigma(k)|^2
-2C_3\sqrt{N}\phi_0 \sigma(k=0)-\frac{\phi_0^2 NV}{\lambda} \nonumber\\
&& +{\rm ~classical~part}.
\label{lea}
\end{eqnarray}
Details of the calculation leading to \eq{lea} as well as the
calculations of the coefficients $C_n~(n=0,1,2,3)$ have been given in
Appendix B. We have retained terms only up to ${\cal O}(1)$ in large
$N$ and up to two derivatives of $\sigma(x)$. The ``classical part''
refers to the contribution which comes from the action evaluated on
the classical solution.

Notice that even though the scalar field started out as an auxiliary
field, it has now developed kinetic terms. As can be seen from
equations \eq{c1}-\eq{dimless}, these terms imply a maximum attainable
velocity at low energies for the $\sigma$ particle, which is given by
\bea
c_\sigma^2=\frac{C_2}{C_1}=c_\psi^2 \ \frac{\bar{C_2}}{\bar{C_1}}, 
\label{LIV}
\eea 
where
$c_\psi=|g_1|$ is the maximum attainable velocity of the fermions at
low energies. The quantities ${\bar C_{1,2}}$ have been defined in
\eq{dimless}. In general $c_\sigma \neq c_\psi$, leading to LIV at
low energies. As a check on our calculations, it is not difficult to
see that for $g_0=0$, $c_\sigma=|g_1|$. Thus, we recover properties of
the usual relativistic NJL model in the appropriate limit. Note that
for $g_0 \neq 0$, the model is renormalizable since the momentum
integrals involved in the computation of $C_{1,2}$ are finite in the
limit $\L \rightarrow \infty$. However, for $g_0=0$, a finite $\L$
is essential since in this usual, nonrenormalizable, relativistic NJL
case the momentum integrals diverge.

What is the extent of the LIV at low energies? It follows from the
above discussion that the magnitude of LIV depends on the deviation of
the ratio $c_\sigma/c_\psi$ from unity. From equations
\eq{c1}-\eq{dimless} in Appendix B, we see that this quantity depends on
the various energy scales only through the dimensionless
ratios\footnote{The reason for the subscript ``R'' will be clear in
the next section where we discuss the renormalization of the
various couplings.}
\bea
\mu_R \equiv \mu/E, \qquad \bar{\mu} \equiv \mu/\L. 
\label{muel}
\eea
In Figure
\ref{fig} we have plotted $c_\sigma/c_\psi$ as a function of $\mu_R^{-1}$ for
different values of $\bar{\mu}$. 
\begin{figure}[htb] 
\centering 
\includegraphics[height=5cm,
width=8cm]{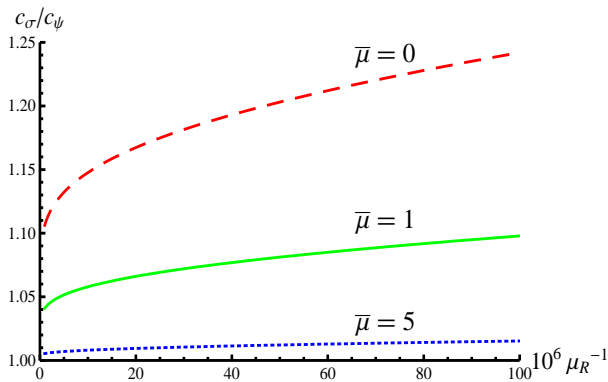}
\caption{The variation of the ratio $c_\sigma/c_\psi$,
or equivalently $({\bar C_2}/{\bar C_1})^{1/2}$, {\em cf.}  \eq{LIV},
with $\mu_R^{-1}$ for different values of $\bar{\mu}$. This ratio
measures Lorentz invariance violations. The dashed curve corresponds to
$\bar{\mu}=0$. The other two cases have $\bar{\mu}=1$ and $5$. We have used
$\phi_0/\mu=10^{-3}$.}
\label{fig}
\end{figure}
In these plots, we have chosen the relative sign of $g_0$ and $g_1$ to
be positive, i.e. $\epsilon=+1$.  The dashed line in Figure
\ref{fig} corresponds to case with $\bar{\mu}=0$, i.e. the case in 
which the cut-off $\L$ is infinite. We see that even at very large
values of $\mu_R$ there is violation of Lorentz invariance at ${\cal
O}(1)$ level. The origin of these LIVs becomes clear from the other
two lines in Figure \ref{fig}. The full line corresponds to the case
with $\bar{\mu}=1$ and for the dotted line $\bar{\mu}=5$. We see that
the LIVs are dramatically reduced in these cases, being smaller for
larger value of $\bar{\mu}$. Clearly, fermion modes with energy larger
than $\mu$ propagating in the loop (Eq. \eq{log-det}) make Lorentz
violating contributions to $C_{1,2}$ at ${\cal O}(1)$ level. These
modes can be removed from the loop by imposing a cut-off which is
smaller than the LIV scale, i.e. for $\bar{\mu} > 1$. For smaller
cut-off, the effect should be better. This is precisely what we see in
the calculations shown in Figure \ref{fig}.

It should be emphasized that the ${\cal O}(1)$ violations of Lorentz
invariance at low energies that we have found in the bosonic sector of
the present model cannot be tuned away by adjusting any parameter.
This makes (an appropriately gauged version of) the present model an
unsuitable candidate for an alternative solution to the hierarchy
problem. A way out could be provided by a supersymmetric extension of
the model. If supersymmetry is broken at a scale $M_s$ much smaller
than the Lorentz invariance violating scale $\mu$, then a cancellation
with bosonic partners would remove the Lorentz violating contributions
from fermionic modes with energy larger than $M_s$ propagating in the
loop. This essentially means that the role of the cut-off $\L$ would
then be played by $M_s$.  The residual low energy LIV would then be
controlled by $M_s^2/\mu^2$, which can be made small by tuning the
scale at which supersymmetry is broken. Existing constraints on LIV
coming from bounds on the maximum attainable velocity of various
particles ($\delta c < 10^{-24}$)
\cite{Liberati:2009pf}-\cite{Scully:2008jp} implies $M_s/\mu < 10^{-12}$.
For $\mu \sim 10^{17}$GeV, this means that supersymmetry can be broken
at a much higher scale, $M_s \sim 100$ TeV, than in the currently
popular scenarios\footnote{For a recent review, see
\cite{Altarelli:2009bz}.}. The present scenario might become a serious
possibility, should LHC see a composite Higgs but no supersymmetry.

\section{RG flows and fixed points}

Classically, at high enough energies (where all masses can be
ignored), the action in \eq{simplestaction} has the scaling symmetry
under which $x^0 \rightarrow x^0/a^3, \ \vec{x} \rightarrow \vec{x}/a,
\ \Psi \rightarrow a^{3/2}\Psi, \ \sigma \rightarrow a^{3}\sigma$,
where $a$ is the parameter of scaling transformation. This is the
$z=3$ Lifshitz-like fixed point. For $g_0=0$, classically the free
fermi theory has the scaling symmetry $x^0 \rightarrow x^0/a, \ \vec{x}
\rightarrow \vec{x}/a, \ \Psi \rightarrow a^{3/2}\Psi$. This is the 
familiar Lorentz invariant case, which can be described in the above
language as a $z=1$ fixed point. At energies much smaller than the
scale $\mu$, Lorentz violations are small even for $z=3$, and so
classically one recovers an approximately Lorentz invariant theory at
low energies. In the following, we will study RG flow in the quantum
theory from the $z=3$ fixed point in the ultraviolet to find out what
theory it flows to in the infrared.

\subsection{Determination of renormalized parameters}

In order to study the flow from high to low energies, we need to find
out how the various couplings get renormalized. The starting point in
the determination of the renormalization of the couplings in action
\eq{momaction}, in the leading large $N$ approximation, is the low
energy action \eq{lea}. To implement the Wilson RG
procedure, we need to rescale the cut-off $E$ in
\eq{lea} back to the original cut-off $\L$. As discussed
above equation \eq{masslv}, our cut-off procedure imposes the
restriction $(k_4^2+l^2(k)) \leq \L^2$ on Euclidean space momentum
integrals. Writing $E=b\L$, we see that the cut-off $E$ on the low
energy action \eq{lea} can be rescaled to $\L$ by the scale
transformations (change of variables) $k_4 \rightarrow b k_4,\ k
\rightarrow a k$, followed by the scalings of the couplings
\bea
a^3 g_0=b g_0', \qquad  a g_1=b g_1'. 
\label{grg2}
\eea
These give the renormalized couplings and RG flow in the free
fermion theory.  Notice that the scaling parameter, $b$, for the
energy is, a priori, unrelated to the scaling parameter, $a$, for the
momenta. The $z=3$ fixed point behaviour corresponds to choosing
$a=b^{1/3}$, and then the couplings scale as $g_0'=g_0$ and
$g_1'=b^{-2/3}g_1$. Since in this case $g_0$ is invariant under the RG
flow, we can set it to unity by scaling $k
\rightarrow |g_0|^{-1/3} k$, leaving only one independent coupling, 
namely $g_1$. Choosing $a=b$ instead, one gets $g_0'=b^{2/3}g_0$ and
$g_1'=g_1$, which are the scalings appropriate for a $z=1$ fixed point.
In this case, $g_1$ is invariant under the RG flow and so we can
scale it away by $k \rightarrow k/|g_1|$. Once again we are left with
only one independent coupling.

The two fixed point behaviours discussed above can be treated together
by setting $a=b^{1/z}$ where $z=3$ or $1$ \footnote{Away from the fixed
points, in general the RG equations \eq{grg2} describe a flow in two
parameters, namely $a$ and $b$. More generally, in a theory with
anisotropy in $n$ different directions, the RG equations will describe
an $n$-parameter flow. It would be interesting to explore such more
general flows. Here we will confine ourselves to a more traditional
view of RG as a flow in a single scale parameter.}. Then, using the
cut-off $\L$ to define the dimensionless renormalized couplings, we
get
\bea
g_{0R} \equiv \L^{\frac{3}{z}-1}g_0'=E^{\frac{3}{z}-1}g_0, \qquad 
g_{1R} \equiv \L^{\frac{1}{z}-1}g_1'=E^{\frac{1}{z}-1}g_1.
\label{dimlessgrg}
\eea
They satisfy the RG equations
\bea
\dot{g}_{0R}=-\left(\frac{3}{z}-1\right)g_{0R}, \qquad  
\dot{g}_{1R}=-\left(\frac{1}{z}-1\right) g_{1R},
\label{grg}
\eea
where a dot denotes a derivative with respect to ($-{\rm
ln}E)$ \footnote{Note that in this convention, the RG flow is from high
to low energies. This is opposite to the convention generally used in
high energy physics.}. Using these two couplings, we can define the
renormalized version of
\eq{masslv}:
\bea
|g_{1R}^3/g_{0R}|^{\frac{1}{2}}=\mu/E=\mu_R.
\label{mur}
\eea
This is precisely the quantity defined in \eq{muel}.

In the leading large-$N$ approximation, the free field renormalization
\eq{grg} is not affected by the Yukawa
coupling. However, the 't Hooft coupling $\lambda$ does receive
quantum corrections. Its renormalization can be deduced from the term
proportional to $C_0$ in the low energy action
\eq{lea}. Scaling the cut-off $E$ back to $\L$ 
in \eq{lea} and using the expression for $C_0$ 
given in \eq{c0}, we get 
\bea
\frac{1}{\lambda'}=\frac{b}{a^3}\biggl\{\frac{1}{\lambda}-
2\int \frac{[d^4k]_{E \rightarrow \L}}{(2\pi)^4} \
\frac{k_4^2+l^2(k)-\phi_0^2}{(k_4^2+l^2(k)+\phi_0^2)^2} \biggr\}, 
\label{lrg0}
\eea
Now, using $(\int [d^4k]_{E
\rightarrow \L} \ \cdots)=(\int [d^4k]_\L \ \cdots)-(\int [d^4k]_E \ \cdots)$
and substituting
$a=b^{1/z}$, we can simplify this equation to get
\bea
\frac{1}{\lambda'}=b^{1-\frac{3}{z}} \biggl\{\xi+
2\int \frac{[d^4k]_E}{(2\pi)^4} \
\frac{k_4^2+l^2(k)-\phi_0^2}{(k_4^2+l^2(k)+\phi_0^2)^2} \biggr\},
\label{lrg1}
\eea 
where\footnote{In the broken phase, $\xi=0$ by the gap equation
\eq{gapeqn}. In the unbroken phase, $\xi$ is a non-zero constant, 
independent of $E$. For this reason, it turns out that the RG equation
obtained in \eq{lrg3}, for the dimensionless coupling $\lambda_R$
defined in \eq{lrg2}, does not depend on $\xi$. Consequently the RG
equation in the unbroken phase can be obtained from \eq{lrg3} by
specializing to $\phi_R=0$.}
\bea
\xi \equiv \frac{1}{\lambda}-2\int \frac{[d^4k]_\L}{(2\pi)^4} \
\frac{k_4^2+l^2(k)-\phi_0^2}{(k_4^2+l^2(k)+\phi_0^2)^2}.
\label{xi}
\eea
So, for the dimensionless renormalized coupling, $\lambda_R \equiv
\L^{\frac{3}{z}-1}\lambda'$, we get
\bea
\frac{1}{\lambda_R}
=\frac{1}{E^{\frac{3}{z}-1}} \biggl\{\xi+2\int \frac{[d^4k]_E}{(2\pi)^4} \ 
\frac{k_4^2+l^2(k)-\phi_0^2}{(k_4^2+l^2(k)+\phi_0^2)^2} \biggr\}.
\label{lrg2}
\eea 
This leads to the RG equation  
\bea
\dot{\lambda}_R=-\left(\frac{3}{z}-1\right)\lambda_R+
\frac{(1-\phi_R^2)}{(1+\phi_R^2)^2} \ \frac{\lambda_R^2}{\pi^3 |g_{0R}|}
\int_0^{h_0} ds \ \frac{s^2}{\sqrt{1-s^2(\epsilon s^2+\mu_R^{\frac{2}{3}})^2}}
\label{lrg3}
\eea
where $\epsilon=\pm 1$ is the relative sign of $g_0$ and $g_1$,  
$\phi_R \equiv \phi_0/E$ is the dimensionless renormalized
coupling corresponding to the vev (with the RG equation
$\dot{\phi}_R=\phi_R$) and
\bea
h_0=\left(\frac{1}{2}+\sqrt{\frac{1}{4}+
\frac{\mu_R^2}{27}}\right)^{\frac{1}{3}}-
\frac{\mu_R^{\frac{2}{3}}}{3}
\left(\frac{1}{2}+\sqrt{\frac{1}{4}+\frac{\mu_R^2}{27}}\right)^{-\frac{1}{3}}
\label{h0}
\eea
Note that the form in which the right-hand side of \eq{lrg3} has been
written is inappropriate for the special case $g_{0R}=0$. In this case
one must use the alternative, but entirely equivalent, form:
\bea
\dot{\lambda}_R=-\left(\frac{3}{z}-1\right)\lambda_R+
\frac{(1-\phi_R^2)}{(1+\phi_R^2)^2} \ \frac{\lambda_R^2}{\pi^3 |g_{1R}|^3}
\int_0^{h_1} ds \ \frac{s^2}{\sqrt{1-s^2(\epsilon \mu_R^{-2}s^2+1)^2}},
\label{lrg4}
\eea
where $h_1=h_0\mu_R^{\frac{2}{3}}$. It is easy to see that $h_1
\rightarrow 1$ as $\mu_R \rightarrow \infty$.

\subsection{The renormalized action}

In terms of the dimensionless renormalized couplings, the low energy
action \eq{lea} can be written as
\bea
S &=& \int \frac{[d^4k]_\L}{(2\pi)^4} \int \frac{[d^4q]_\L}{(2\pi)^4} \
\bar\Psi_{iR}(k) \biggl\{(2\pi)^4 \L \biggl(\tilde{k}_0 \gamma^0-
\vec{l_R}(\vec{\tilde{k}}).\vec{\gamma}
+\phi_R\biggr)\delta^4(k-q) \nonumber \\
&& +\frac{1}{\sqrt{N}}\sigma_R(k-q)\biggr\}
\Psi_{iR}(q)-\L^{\frac{3}{z}-1} \int \frac{[d^4k]_\L}{(2\pi)^4} 
\left(\frac{1}{\lambda_R}+C_{1R}\tilde{k}_0^2-C_{2R}\tilde{k}^2 \right)
|\sigma_R(k)|^2 \nonumber \\
&& -2\frac{\L^{\frac{3}{z}}}{\lambda_R}\sqrt{N}\phi_R \sigma_R(k=0) \ 
+ \ \text{classical part},
\label{rlea}
\eea
where $\vec{l_R}(\vec{\tilde{k}})=\vec{\tilde{k}}
(g_{0R}\tilde{k}^2+g_{1R}),\ \tilde{k}_0=k_0/\L,\
\tilde{k}=k\L^{-\frac{1}{z}}$. Moreover, the renormalized fields are related
to the bare fields by $\Psi_{iR}=b^{\frac{3}{2z}+1}\Psi_i,\
\sigma_R=b^{\frac{3}{z}}\sigma$. The coefficients $C_{1,2R}$ are related to 
$C_{1,2}$ and have been defined in \eq{rdimless2}.

This action seems to depend separately on the two couplings $g_{0R},\
g_{1R}$, but actually the physics described by it depends only on the
combination $\mu_R$, \eq{mur}. To see this more explicitly, let us
make the change of variables $\tilde{k} \rightarrow
\tilde{k}/|g_{1R}|,\ \Psi_{iR} \rightarrow |g_{1R}|^{\frac{3}{2}}\Psi_{iR},\ 
\sigma_R \rightarrow |g_{1R}|^{3}\sigma_R,\
\lambda_R \rightarrow \lambda_1=\lambda_R/|g_{1R}|^{3}$. 
After this change of variables, the action takes the form
\bea
S &=& \int \frac{[d^4k]_\L}{(2\pi)^4} \int \frac{[d^4q]_\L}{(2\pi)^4} \
\bar\Psi_{iR}(k) \biggl\{(2\pi)^4 \L \biggl(\tilde{k}_0 \gamma^0-
\vec{\tilde{l}}_R(\vec{\tilde{k}}).\vec{\gamma}
+\phi_R\biggr)\delta^4(k-q) \nonumber \\
&& +\frac{1}{\sqrt{N}}\sigma_R(k-q)\biggr\}
\Psi_{iR}(q)-\L^{\frac{3}{z}-1} \int \frac{[d^4k]_\L}{(2\pi)^4} 
\left(\frac{1}{\lambda_1}+\bar{C}_1\tilde{k}_0^2-
\bar{C}_2\tilde{k}^2 \right)|\sigma_R(k)|^2 \nonumber \\
&& -2\frac{\L^{\frac{3}{z}}}{\lambda_1}\sqrt{N}\phi_R \sigma_R(k=0) \ 
+ \ \text{classical part},
\label{rlea1}
\eea
where $\vec{\tilde{l}}_R(\vec{\tilde{k}})=
\vec{\tilde{k}} (\epsilon \tilde{k}^2/\mu_R^2+1)$ 
and $\bar{C}_{1,2}$
are given by \eq{rdimless1}. This form of the action makes it
explicit that physics depends only on the combination $\mu_R$ since
any separate dependence on $g_{0R},\ g_{1R}$ has now disappeared.

The form \eq{rlea1} of the low energy action is not suitable
for small values of $\mu_R$ (equivalently for small values of
$g_{1R}$ or large values of $g_{0R}$). In this case, a more suitable
change of variables in the action \eq{rlea} is $\tilde{k} \rightarrow
\tilde{k}|g_{0R}|^{-\frac{1}{3}},\ 
\Psi_{iR} \rightarrow |g_{0R}|^{\frac{1}{2}}\Psi_{iR},\ 
\sigma_R \rightarrow |g_{0R}|\sigma_R,\
\lambda_R \rightarrow \lambda_3=\lambda_R/|g_{0R}|$.
After this change of variables, the action takes the form
\bea
S &=& \int \frac{[d^4k]_\L}{(2\pi)^4} \int \frac{[d^4q]_\L}{(2\pi)^4} \
\bar\Psi_{iR}(k) \biggl\{(2\pi)^4 \L \biggl(\tilde{k}_0 \gamma^0-
\vec{\tilde{l}}_R'(\vec{\tilde{k}}).\vec{\gamma}
+\phi_R\biggr)\delta^4(k-q) \nonumber \\
&& +\frac{1}{\sqrt{N}}\sigma_R(k-q)\biggr\}
\Psi_{iR}(q)-\L^{\frac{3}{z}-1} \int \frac{[d^4k]_\L}{(2\pi)^4} 
\left(\frac{1}{\lambda_3}+C_1'\tilde{k}_0^2-C_2'\tilde{k}^2 \right)
|\sigma_R(k)|^2 \nonumber \\
&& -2\frac{\L^{\frac{3}{z}}}{\lambda_3}\sqrt{N}\phi_R \sigma_R(k=0) \ 
+ \ \text{classical part},
\label{rlea2}
\eea
where $\vec{\tilde{l}}_R'(\vec{\tilde{k}})=
\vec{\tilde{k}} (\epsilon \tilde{k}^2+\mu_R^{\frac{2}{3}})$ 
and $C_1'=\bar{C}_1\mu_R^{-2}, \ C_2'=\bar{C}_2\mu_R^{-2/3}$. It can
be shown that $C_{1,2}'$ have a finite limit as $\mu_R \rightarrow 0$;
see equations \eq{dimless}-\eq{dimless3}.  This form of the low energy
action is now suitable for small values of $\mu_R$.

We have thus found two equally valid descriptions of the physics of
the 4-fermi theory. One is that given by the action in
\eq{rlea}, which is valid for all the values of the renormalized 
coupling $\mu_R$. However, in this form the action depends
on two couplings, $g_{0R}$ and $g_{1R}$.  In the form \eq{rlea1} and
\eq{rlea2}, the low energy action depends only on the combination 
$\mu_R$ of these, but two different descriptions are needed to cover the
entire range of possible values of $\mu_R$.

\subsection{Fixed points}

As we have argued above, the relevant renormalized coupling constants
in the low energy theory are $\lambda_1=\lambda_R/|g_{1R}|^{3}$ and
$\lambda_3=\lambda_R/|g_{0R}|$, with $\lambda_3=\mu_R^2 \lambda_1$.
The RG equations for these can be obtained from equations \eq{lrg3}
and \eq{lrg4} using \eq{grg}. We get
\bea
\dot{\lambda}_3 &=& \frac{(1-\phi_R^2)}{(1+\phi_R^2)^2} \ 
\frac{\lambda_3^2}{\pi^3}
\int_0^{h_0} ds \ \frac{s^2}{\sqrt{1-s^2(\epsilon s^2+\mu_R^{\frac{2}{3}})^2}}.
\label{lrg31} \\
\dot{\lambda}_1 &=& -2 \lambda_1+
\frac{(1-\phi_R^2)}{(1+\phi_R^2)^2} \ 
\frac{\lambda_1^2}{\pi^3}
\int_0^{h_1} ds \ \frac{s^2}{\sqrt{1-s^2(\epsilon \mu_R^{-2}s^2+1)^2}}, 
\label{lrg41} 
\eea
Together with these, we also have the RG equation for $\mu_R$, namely
\bea
\dot{\mu_R}=\mu_R, \qquad ({\mu_R^{-1}})^{\hbox{$\cdot$}} =\mu_R^{-1}.
\label{mur1}
\eea 
The second of these is appropriate for large $\mu_R$. Equations
\eq{lrg31}-\eq{mur1} constitute the set that describes the RG flows in this
model\footnote{Note that $\phi_R$ is not an independent variable since
it is determined in terms of $\lambda_{1,3}$ and $\mu_R$ by the gap
equation in the broken phase, while in the unbroken phase it
vanishes.}. We emphasize that the explicit dependence on $z$ has
dropped out of these equations. This is nice since one expects that
specific values of $z$ should characterize only the end points of an
RG trajectory, not the trajectory itself.

Now, let us first consider the case of small $\mu_R$. In this case,
the appropriate equation is \eq{lrg31}. We see that there is a
possible fixed point at $\lambda_3=0$. For this to be a fixed
point, we must also have $\mu_R=0$ and $\phi_R=0$. This is what we have
been describing as the $z=3$ Lifshitz-like fixed point.

The case $\mu_R \rightarrow \infty$ is more interesting. In this case
we must use \eq{lrg41}, which in the limit approximates to the
equation
\bea
\dot{\lambda}_1=-2\lambda_1+
\frac{(1-\phi_R^2)}{4\pi^2(1+\phi_R^2)^2}\lambda_1^2.
\label{l3rg}
\eea 
For a fixed point we must have $\phi_R=0$ and one of the two
possibilities: $\lambda_1=0, \ 8\pi^2$. The first of these is
the free field (Gaussian) fixed point and the second is a new Lorentz
invariance violating fixed point. 

Figure \ref{fig2} shows a plot of
the RG flows near the three fixed points we have found. The data for
this figure have been obtained using the exact RG equations \eq{lrg31} and
\eq{lrg41}. Note that we have used $\epsilon=+1$ in these calculations.
\begin{figure}[htb] 
\centering 
\includegraphics[height=6cm,
width=6.5cm]{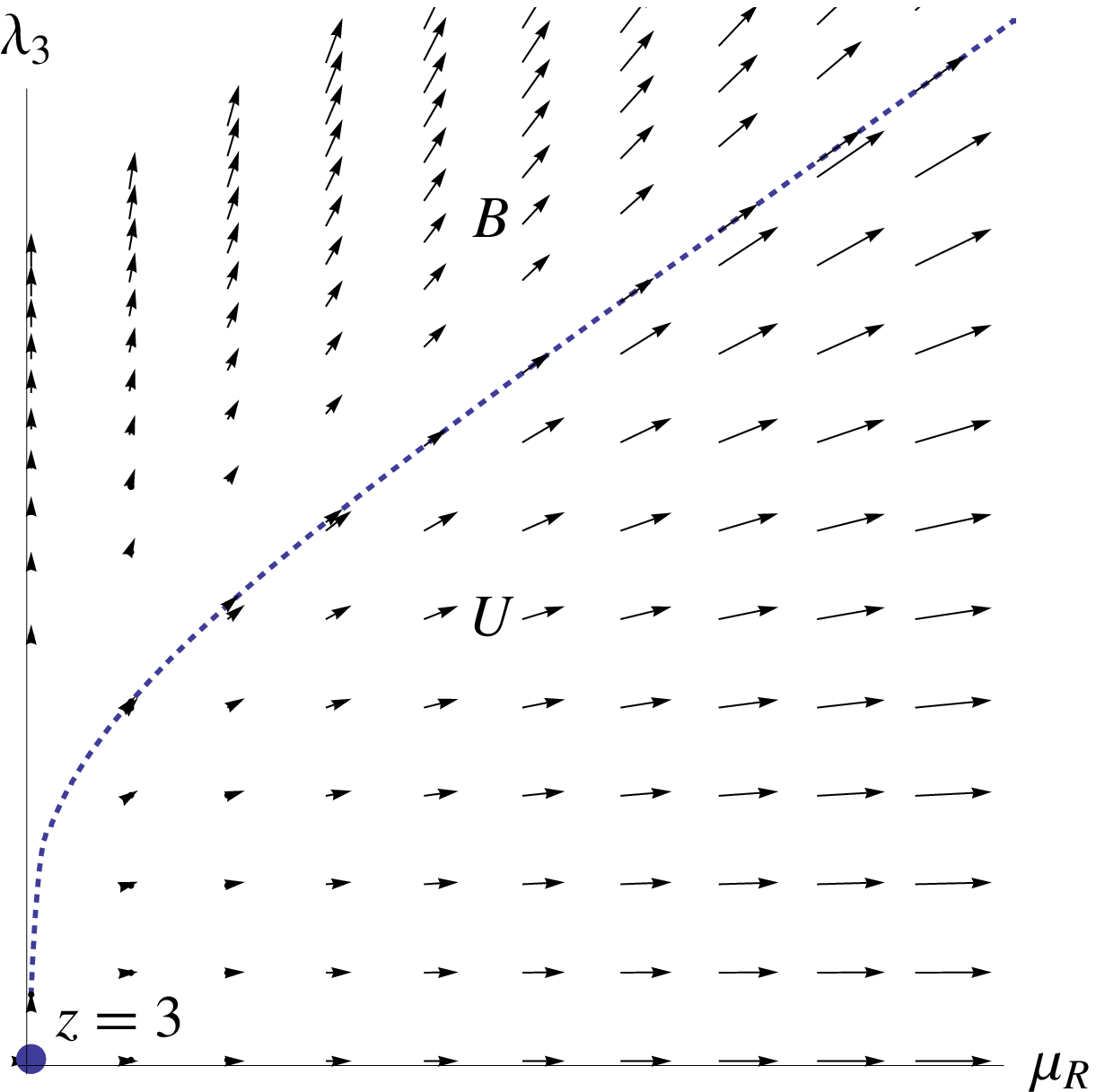}~~\includegraphics[height=6cm, width=6.5cm]{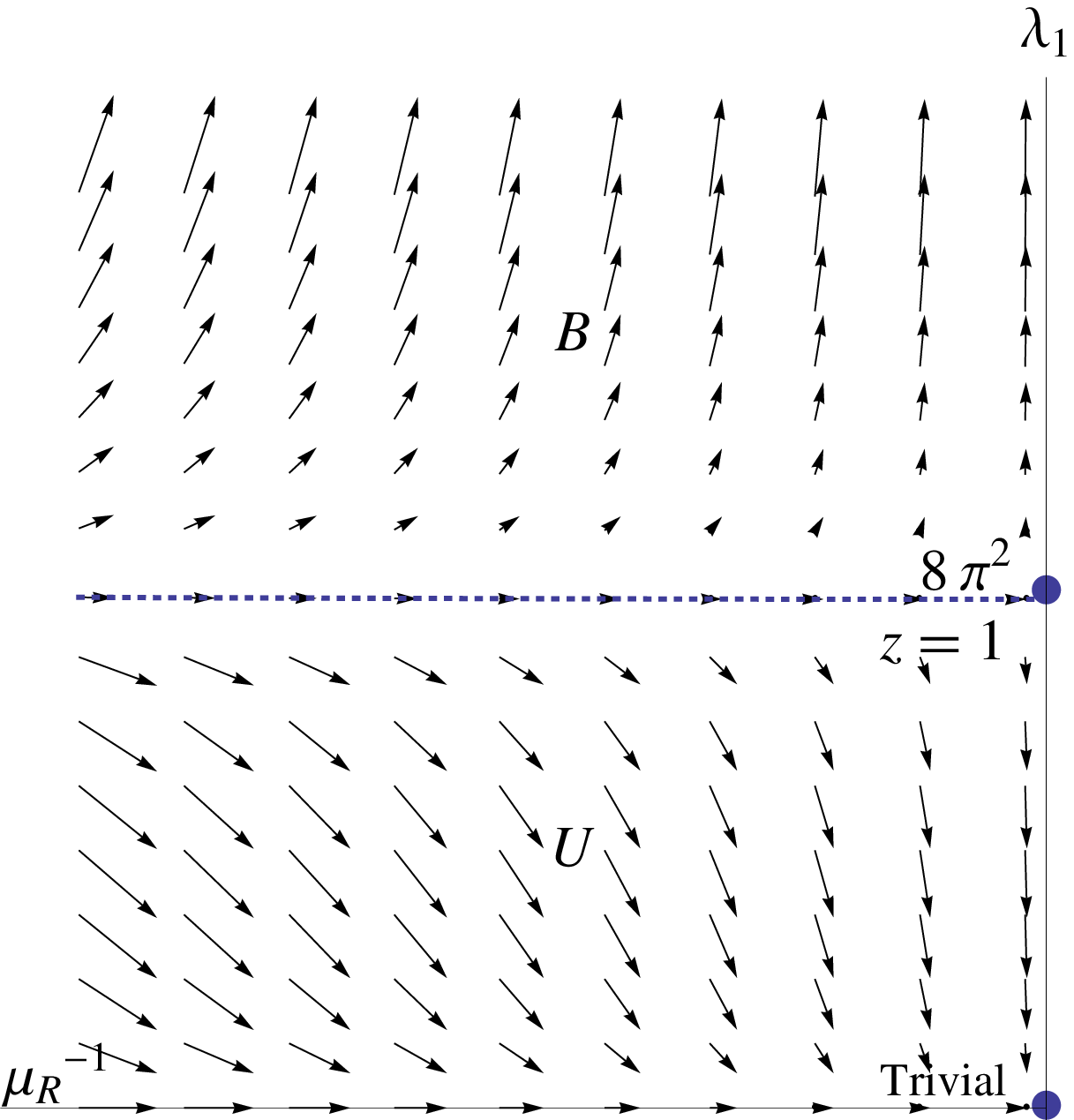}
\caption{In the figure on the left are plotted RG flows in the 
$(\lambda_3,\ \mu_R)$ plane and in the one on the right are
flows in the $(\lambda_1,\ 1/\mu_R)$ plane. $\mu_R$
increases from left to right in both figures. The chiral symmetry
broken (unbroken) phase is indicated by the letter B (U). The dotted
line is the critical curve on which $\phi_0$ vanishes.}
\label{fig2}
\end{figure}
In the broken phase, on the critical line $\phi_R=0$, the RG flow from
UV ends on the Lorentz invariance violating fixed point in the
IR. This can be seen directly from \eq{lrg2}. Making the change of
variables $k_4
\rightarrow Ek_4,\ k \rightarrow Ek/|g_1|$ in this equation, we get
\bea
\frac{1}{\lambda_1}=2\int \frac{[d^4k]_1}{(2\pi)^4} \ 
\frac{k_4^2+\tilde{l}_R^2(k)-\phi_R^2}
{(k_4^2+\tilde{l}_R^2(k)+\phi_R^2)^2}+\frac{|g_1|^3}{E^2}\xi,
\label{rgsol}
\eea
where, as before, $\tilde{l}_R(k)=k(\epsilon k^2/\mu_R^2+1)$ and 
$\int [d^4k]_1 \ \cdots=\int d^4k \ 
\theta(1-(k_4^2+\tilde{l}_R^2(k))) \ \cdots$. Now, in the 
broken phase, the gap equation implies $\xi=0$. So, for
$\mu_R^{-1}=\phi_R=0$, the right-hand side of the above equation
evaluates to $1/8\pi^2$. What happens for $\mu_R^{-1}=0$, but $\phi_R
\neq 0$? In this case, for small values of $\phi_R$, the coupling 
increases as $\lambda_1 \sim 8\pi^2/(1-\phi_R^2 {\rm
ln}\phi_R^{-2})$. So, these trajectories diverge to larger values of
the coupling, doing so faster for larger values of $\phi_R$. Figure 2
confirms this for trajectories in the broken phase. Note that there is
no fixed point for $\lambda_1 \rightarrow
\infty$ since the beta function of $\lambda_1^{-1}$ does not vanish at
$\lambda_1^{-1}=0$.

In the unbroken phase, $|g_1|^3 \xi$ is a non-zero constant,
independent of the flow parameter $E$, while $E \rightarrow 0$ in the
IR. Thus, in the unbroken phase the RG flow will terminate at
$\lambda_1=0$ in the IR.

What does the theory look like at these two fixed points? Consider
first the nontrivial fixed point at $\lambda_1=8\pi^2$. The
fixed point action can be obtained from \eq{rlea1} by setting
$\phi_R=0$ and taking the limit $\mu_R \rightarrow \infty$. For
$\phi_R=0$, the fermions become massless. In the $\sigma$ kinetic
terms, the coefficients $\bar{C}_{1,2}$ grow logarithmically in the
limit $\mu_R \rightarrow \infty$, as shown in
\eq{c12r1}. The implication is that as we approach the fixed point  
$\lambda_1=8\pi^2$, the kinetic terms for $\sigma$ grow and
eventually dominate the mass term. This can be seen more directly by
the rescaling $\sigma \rightarrow \sigma/\sqrt{\bar{C}_1}$. In the limit
$\mu_R \rightarrow \infty$, the mass term and the Yukawa interaction
disappear, leaving behind a massless scalar decoupled from the
fermions. So the theory at this fixed point has free massless fermions
and a free massless scalar, the maximum attainable velocity of the
latter being different from the former, unless $\bar{\mu} \gg 1$, in
which case Lorentz invariance is restored near the fixed point.

Note that our analysis implies the existence of a fixed point in the
usual relativistic NJL model as well. This can be established as
follows. The RG equation for the 4-fermi coupling in the NJL model can
be obtained by setting $\bar{\mu} \gg 1$ and $z=1$ in
\eq{lrg4}. Since $\mu_R \gg \bar{\mu}$, this also implies $\mu_R \gg
1$. For large $\mu_R$ and $z=1$, the equation for
$\lambda_1=\lambda_R/|g_{1R}|^3$ is precisely
\eq{l3rg}. Moreover, from the low energy action \eq{rlea1}, we see
that in this parameter regime the Lorentz violating piece in the
fermion kinetic term vanishes and the coefficients $\bar{C}_{1,2}$
work out to be those appropriate for a relativistic NJL model with a
cut-off $\L$, as we have argued below equation \eq{c12r1}.

The other fixed point, that at $\lambda_1=0$, is described by just a
free massless fermion. This is because near this fixed point the
$\sigma$ mass goes to infinity as $\mu_R^2$ because of the manner in
which $\lambda_1$ approaches the fixed point, which is described by
equation \eq{rgsol}. Therefore, this time the rescaling $\sigma
\rightarrow \sigma/\sqrt{\bar{C}_1}$ leaves the mass term as dominant,
with the mass going to infinity. Hence $\sigma$ decouples at the fixed
point, leaving behind free massless relativistic fermions. This is the
theory that one gets from the original four-fermi model at the trivial
(Gaussian) fixed point.

\section{Concluding remarks}

In this paper we have analysed RG flows in a $z=3$ Lifshitz-like four
fermi model, which is ultraviolet complete in $3+1$ dimensions. The
model flows in the infrared to a theory in which Lorentz invariance is
violated at ${\cal O}(1)$ level, which cannot be tuned away by adjusting a
parameter. The origin of these violations can be traced to fermions
with energies higher than the Lorentz violating energy scale,
propagating in loops and contributing to the induced kinetic terms for
the composite boson, in the chiral symmetry broken vacuum. However, if
one works with a finite cut-off, which is taken to be much smaller
than the Lorentz invariance violating scale, then the model flows in
the infrared to an approximately Lorentz invariant theory even in the
bosonic sector, which is similar to the low energy limit of the usual
Nambu$-$Jona-Lasinio model in the broken phase. A physical way of
interpreting the cut-off could be as supersymmetry breaking scale
in a supersymmetric version of this model. In this case, the offending
contributions of fermions in loops would be cancelled by their
supersymmetric partners and the Lorentz violations would be controlled
by the ratio of the Lorentz violating scale to the supersymmetry
breaking scale, which can, in principle, be made small.  Possible
applications of the present model to the Higgs sector of the Standard
Model would then put constraints on these two scales for consistency
with data.

A remarkable feature of the general RG equations we have obtained,
\eq{grg2} and \eq{lrg0}, is that they describe flow in two
scaling parameters, namely $a$ and $b$. More generally, in a theory
with anisotropy in $n$ different directions, the RG equations will
describe flow in $n$ parameters. The parameters presumably get related
near a fixed point, as in the present example in which we found that
$a=b^{\frac{1}{z}}$ near the fixed point labeled by the exponent
$z$. Away from the fixed points, however, a more general flow in
multiple scaling parameters would seem to be more appropriate. It
would be interesting to explore such more general flows.

\section{Acknowledgements}

We would like to thank Spenta Wadia for a collaboration at an early
stage of this work and for numerous discussions.  We would also like
to thank Sumit Das, Alfred Shapere, Juan Maldacena, Shiraz Minwalla
and Michael Peskin for discussions.  G.M. would like to thank the
organizers of the Benasque conference on Gravity (July 2009), the
organizers of the QTS6 meeting in Lexington, the University of
Kentucky, Lexington and the School of Natural Sciences, IAS, Princeton
for hospitality during part of this project.

\appendix

\section{Coefficients appearing in the vacuum solution}

We give here the coefficients $I_0,~I_1,~I_2$ and $I_3$ that appeared
in equation \eq{rho-0}:
\begin{eqnarray}
\mm I_0\equiv \int_0^1 dk \frac{k^4}{1+g_0^2k^6}=
\frac{1}{12g_0^{5/3}} \biggl(2 \tan^{-1}{\frac{3 g_0^{1/3}(1-g_0^{2/3})}
{1+g_0^{4/3}-4 g_0^{2/3}}}+
\sqrt{3}\ln {\frac{1+g_0^{2/3}-\sqrt{3} g_0^{1/3}}{1+g_0^{2/3}+
\sqrt{3} g_0^{1/3}}}\biggr),\nonumber\\
\mm I_1\equiv \int_0^1 dk \frac{k^2}{1+g_0^2k^6}
=\frac{1}{3g_0} \tan^{-1}{g_0},\nonumber\\
\mm I_2\equiv \int_0^1 dk \frac{k^8}{(1+g_0^2k^6)^2}
=\frac{1}{6g_0^3}\left(\tan^{-1}{g_0}-\frac{g_0}{1+g_0^2}\right),
\nonumber\\
\mm I_3\equiv \int_0^1 dk \frac{k^8}{(1+g_0^2k^6)^3}
=\frac{1}{24g_0^3}\left(\tan^{-1}{g_0}+ 
\frac{g_0^3-g_0}{(1+g_0^2)^2}\right).
\end{eqnarray}

\section{Evaluation of the low energy effective action}

Here we give some details of the calculation that lead to the low
energy action \eq{lea}. Integrating out the high energy modes of the
fermions gives the effective action 
\begin{eqnarray}
\Delta S=-i~{\rm Tr}~\ln (A+B)=
-i~{\rm Tr}~ \ln(A)-i~{\rm Tr}~\ln (1+A^{-1}B)
\label{log-det}
\end{eqnarray}
where $A(k,q) \equiv (2\pi)^4\left(\gamma^0 k_0-\vec{\gamma}.\vec{l}(\vec{k})
+\phi_0\right)\delta^4(k-q)$, $B(k,q) \equiv
\frac{1}{\sqrt{N}}\sigma(k-q)$ and 'Tr' stands for integration over high 
momenta (between the lower cut-off $E$ and upper cut-off $\L$, as
explained above equation \eq{lea}) and trace over all indices. $\Delta S$
can be expanded as
\begin{eqnarray}
\Delta S=-i~{\rm Tr}~\ln(A)-i~{\rm Tr}(A^{-1}B) +\frac{i}{2}
{\rm Tr}(A^{-1}B)^2+~\text{higher powers}
\end{eqnarray}
Each factor of B comes with a factor of $\frac{1}{\sqrt{N}}$ and hence
the ``higher powers'' of $(A^{-1}B)$ are subleading in $1/N$. Thus,
powers higher than quadratic in $\sigma(k)$ in the effective action
\eqref{lea} are subleading in $1/N$, which we omit. Now,
\begin{eqnarray}
{\rm Tr}~\ln(A)=iNV\int \frac{[d^4k]_{E \rightarrow \L}}{(2\pi)^4} \quad
{\rm tr}~\ln\left(\gamma^0 k_0-\vec{\gamma}.\vec{l}(\vec{k})+\phi_0\right),
\end{eqnarray}
where `tr' stands for trace over Dirac indices only.
Also, ${\rm Tr}(A^{-1}B)$ is given by
\begin{eqnarray}
{\rm Tr} (A^{-1}B)=4 i \phi_0\sqrt{N}\sigma(k=0) 
\int \frac{[d^4k]_{E \rightarrow \L}}{(2\pi)^4} \
\frac{1}{k_{4}^2+l^2(k)+\phi_0^2},
\end{eqnarray}
where we have continued the momenta to the Euclidean signature. Thus,
the coefficient $C_3$ appearing in \eq{lea} is given by
\begin{eqnarray}
C_3 \mm=\frac{1}{\lambda}
-2 \int \frac{[d^4k]_{E \rightarrow \L}}{(2\pi)^4} \ 
\frac{1}{k_4^2+l^2(k)+\phi_0^2}.
\end{eqnarray}
Now,
\begin{eqnarray}
\mm {\rm Tr} (A^{-1}B)^2= \int \frac{[d^4p]_E}{(2\pi)^4} |\sigma(p)|^2
\int \frac{[d^4k]_{E \rightarrow \L}}{(2\pi)^4} \nonumber \\
\mm \times {\rm tr}~ \frac{1}{\left\{\gamma^0 k_0 - 
\vec \gamma.\vec{l}(\vec{k})
+\phi_0\right\}\left\{\gamma^0 (k_0-p_0) 
- \vec \gamma. \vec{l}(\vec{k}-\vec{p})+\phi_0\right\}} \nonumber
\end{eqnarray}
Expanding to quadratic order in $p_0,|\vec{p}|$ for small values, the 
above expression gives
\begin{eqnarray}
\frac{i}{2}{\rm Tr} (A^{-1}B)^2=\mm\int \frac{[d^4p]_E}{(2\pi)^4}|\sigma(p)|^2
\biggl\{C_1 p_0^2-C_2 p^2
+2 \int \frac{[d^4k]_{E \rightarrow \L}}{(2\pi)^4} \
\frac{k_4^2+l^2(k)-\phi_0^2}{\left(k_4^2+l^2(k)+\phi_0^2\right)^2}
\biggr\} \nonumber 
\end{eqnarray}
Thus, the action will be of the form given by equation
\eqref{lea} where $C_0$ is given by
\begin{eqnarray}
C_0=\frac{1}{\lambda}-2\int \frac{[d^4k]_{E \rightarrow \L}}{(2\pi)^4} \
\frac{k_4^2+l^2(k)-\phi_0^2}{\left(k_4^2+l^2(k)+\phi_0^2\right)^2}.
\label{c0}
\end{eqnarray}
Moreover, the coefficients $C_{1,2}$ are given by
\bea
&&C_1(g_0,g_1;\L,E,\phi_0)=\int \frac{[d^4k]_{E \rightarrow \L}}{(2\pi)^4}  
\biggl\{-2 \Delta^2+4\left(l^2(k)+4\phi_0^2\right) \Delta^3
\nonumber \\
&& \hspace{7cm} -16 \phi_0^2\left(l^2(k)+
\phi_0^2\right) \Delta^4 \biggr\},
\label{c1} \\
&&C_2(g_0,g_1;\L,E,\phi_0)=\int \frac{[d^4k]_{E \rightarrow \L}}{(2\pi)^4} 
\biggl\{A_2(k) \Delta^2-A_3(k) \Delta^3
+A_4(k) \Delta^4 \biggr\},\nonumber\\
&&~~~~A_2(k)=2(g_0k^2+g_1)(2 g_0 k^2+g_1)+\frac{4}{3}g_0 k^2 
(5 g_0 k^2+ 3 g_1), \nonumber\\
&&~~~~A_3(k)=\frac{4}{3}l^2(k)(3 g_0 k^2+g_1)^2+
4\phi_0^2 (g_0 k^2+g_1)(3 g_0 k^2+g_1) \nonumber\\
&&~~~~\hspace{1.3cm}+\frac{8}{3} g_0\phi_0^2 k^2(6 g_0 k^2+4 g_1), 
\nonumber\\
&&~~~~A_4(k)=\frac{16}{3} \phi_0^2 l^2(k) (3 g_0 k^2+g_1)^2,
\label{c2}
\eea
where $\Delta(k_4,k)=1/(k_4^2+l^2(k)+\phi_0^2)$. 

For nonzero values of $g_0$ and $g_1$, one can express the dependence
of $C_{1,2}$ on these parameters essentially only through one
combination, the scale $\mu$ defined in \eq{masslv}. For example, one
can scale out the dependence on $g_1$. This can be done by the change
of the integration variable $k \rightarrow k/|g_1|$. We get
\bea
&& C_1(g_0,g_1;\L,E,\phi_0)= |g_1|^{-3}C_1(\epsilon
\mu^{-2},1;\L,E,\phi_0) \equiv |g_1|^{-3}\bar{C}_1,
\nonumber \\
&& C_2(g_0,g_1;\L,E,\phi_0)=
|g_1|^{-1}C_2(\epsilon \mu^{-2},1;\L,E,\phi_0) \equiv |g_1|^{-1}\bar{C}_2.
\label{dimless}
\eea
where $\epsilon=\pm 1$ is the relative sign of $g_0$ and $g_1$. In
terms of the dimensionless renormalized parameters defined in
\eq{muel} and \eq{mur}, it is easy to show that
\bea
\bar{C}_{1,2}=C_{1,2}(\epsilon \mu_R^{-2},1;\mu_R/\bar{\mu},1,\phi_R)=
C_{1,2}(\epsilon,1;1/\bar{\mu},1/\mu_R,\bar{\phi}/\bar{\mu}),
\label{rdimless1}
\eea
where $\bar{\phi}=\phi_0/\L$ and $\bar{\mu}=\mu/\L$ are the values of
the respective running couplings, $\phi_R,\ \mu_R$, at the UV
cut-off. The form in the second equality will be useful later when we
discuss the $E \rightarrow 0$ limit of these coefficients. Now, from the 
first equality of \eq{rdimless1} and \eq{dimless}, we get
\bea
&& \bar{C}_1=|g_{1R}|^3 C_1(g_{0R},g_{1R};\mu_R/\bar{\mu},1,\phi_R) \equiv 
|g_{1R}|^3 C_{1R}, \nonumber \\
&& \bar{C}_2=|g_{1R}| C_2(g_{0R},g_{1R};\mu_R/\bar{\mu},1,\phi_R) \equiv 
|g_{1R}| C_{2R}.
\label{rdimless2}
\eea
Similarly, scaling out the dependence on $g_0$, we get
\bea
&& C_1(g_0,g_1;\L,E,\phi_0)=
|g_0|^{-1}C_1(\epsilon,\mu^{\frac{2}{3}};\L,E,\phi_0),
\nonumber \\
&& C_2(g_0,g_1;\L,E,\phi_0)=
|g_0|^{-1/3}C_2(\epsilon,\mu^{\frac{2}{3}};\L,E,\phi_0).
\label{dimless2}
\eea
Moreover, one can easily show that in terms of the renormalized
parameters,
\bea
&& C_1(\epsilon,\mu^{\frac{2}{3}};\L,E,\phi_0)=E^{-2}
C_1(\epsilon,\mu_R^{\frac{2}{3}};\mu_R/\bar{\mu},1,\phi_R) \equiv C_1'/E^2,
\nonumber \\
&& C_2(\epsilon,\mu^{\frac{2}{3}};\L,E,\phi_0)=E^{-2/3}
C_2(\epsilon,\mu_R^{\frac{2}{3}};\mu_R/\bar{\mu},1,\phi_R) \equiv C_2'/E^{2/3}
\label{dimless21}
\eea
Then, from \eq{rdimless2}-\eq{dimless21}, one gets
\bea
C_1'=\bar{C}_1/\mu_R^2,\qquad C_2'=\bar{C}_2/\mu_R^{\frac{2}{3}}.
\label{dimless3}
\eea 
It is easy to see that $C_{1,2}'$ are finite for $\mu_R=0$. It
follows that the quantities on right-hand side of equations in
\eq{dimless3} have a finite limit as $\mu_R \rightarrow 0$. 

\section{Limiting behaviour of $\bar{C}_{1,2}$}

Here we will discuss the limiting behaviour of the coefficients
$\bar{C}_{1,2R}$, given by \eq{c1}, \eq{c2} and \eq{rdimless1}, for
large values of $\mu_R$, which is the IR regime of small $E$. We first
note that for any positive real number $\eta$ satisfying $\bar{\mu}^{-1}
> \eta^{-1} > \mu_R^{-1}$, we have $(\int [d^4k]_{\mu_R^{-1} \rightarrow 
\bar{\mu}^{-1}} \ \cdots)=(\int [d^4k]_{\eta^{-1} \rightarrow 
\bar{\mu}^{-1}} \ \cdots)-(\int [d^4k]_{\mu_R^{-1} \rightarrow 
\eta^{-1}} \ \cdots)$.
It follows from this identity and the second equality of
\eq{rdimless1} that
\bea
\bar{C}_{1,2}=C_{1,2}(\epsilon,1;1/\bar{\mu},1/\eta,
\bar{\phi}/\bar{\mu})+C_{1,2}(\epsilon,1;1/\eta,1/\mu_R,
\bar{\phi}/\bar{\mu})
\label{c12r}
\eea
We are interested in the limit of large $\mu_R$ for
$\bar{\phi}=0$. The $\mu_R$ dependence of $\bar{C}_{1,2}$ comes
only from the second term above. In the limit $1 \gg \eta^{-1} >
\mu_R^{-1}$, the calculation of this term greatly simplifies since the
``higher derivative'' terms can be neglected. That is, in this regime
of parameter values, $k \ll 1$ throughout the integration range, so we
may neglect $k^2$ compared to $1$. Then, from \eq{c1} and \eq{c2}, we
find that the leading contribution goes as
\bea
\bar{C}_{1,2} \approx C_{1,2}(\epsilon,1;1/\bar{\mu},1/\eta,0)+
\frac{1}{8\pi^2} {\rm ln}\left(\frac{\mu_R}{\eta}\right).
\label{c12r1}
\eea
We see that $\bar{C}_{1,2}$ grow logarithmically with $\mu_R$. If
$\bar{\mu} \gg 1$, then we may choose $\eta=\bar{\mu}$. In this case,
$\bar{C}_1 \approx \bar{C}_2 \approx \frac{1}{8\pi^2} {\rm
ln}\left(\mu_R/\bar{\mu}\right)$. These are just the values for these
coefficients in the relativistic NJL case. However, in general,
$\bar{C}_1$ is quite different from $\bar{C}_2$ and both are quite
different from the corresponding quantities in the relativistic NJL
case. Therefore, in general the $\sigma$ kinetic terms do not have
Lorentz symmetry, as discussed in section 3 below equation \eq{LIV}.

\end{document}